%% file: simplex_solid.tex
\documentclass[twocolumn,showpacs,preprintnumbers,amsmath,amssymb]{revtex4}
\usepackage{dcolumn}
\usepackage{color}
\usepackage{bm}
\usepackage{youngtab}
\usepackage{graphicx}
\input ldefs

\def \be{\begin{equation}}
\def \ee{\end{equation}}
\def \bea{\begin{eqnarray}}
\def \eea{\end{eqnarray}}

\def\xibar{\bar\xi}
\def\zbar{{\bar z}}

\def\bone{{\bf 1}\hskip-5pt{\bf 1}}
\physgreek

\def\Hcl{H\nd_{\rm cl}}
\def\HLT{H\nd_{\rm LT}}
\def\singlet{\hbox to 6pt{$\bullet$\hfill}\kern-1.5pt
{\raise 0.5ex \hbox to 14pt{\leaders\hrule\hfill}}
\kern-1pt\hbox to 6pt{$\bullet$\hfill}}
\def\blank{{\hskip 14pt}}
\def\uar{\uparrow}
\def\dar{\downarrow}

\begin{document}
\title{Simplex solid states of SU($N$) quantum antiferromagnets}
\author{Daniel P. Arovas}
\address{Department of Physics, University of California at San Diego,
La Jolla, CA 92093}

\date{\today}
\begin{abstract}
I define a set of wavefunctions for SU($N$) lattice antiferromagnets, analogous to
the valence bond solid states of Affleck \etal\ \cite{AKLT}, in which the singlets are extended
over $N$-site simplices.  As with the valence bond solids, the new simplex solid (SS)
states are extinguished by certain local projection operators, allowing one to
construct Hamiltonians with local interactions which render the SS states exact
ground states.  Using a coherent state representation, I show that the quantum
correlations in each SS state are calculable as the finite temperature correlations
of an associated classical model, with $N$-spin interactions, on the same lattice. 
In three and higher dimensions, the SS states can spontaneously break SU($N$)
and exhibit $N$-sublattice long-ranged order, as a function of a discrete parameter
which fixes the local representation of SU($N$).  I analyze this transition using
a classical mean field approach.  For $N>2$ the ordered state is selected via
an `order by disorder' mechanism.  As in the AKLT case, the bulk representations
fractionalize at an edge, and the ground state entropy is proportional to the volume
of the boundary.
\end{abstract}
\maketitle
\vskip2pc 
\narrowtext


\Yvcentermath1
\Yautoscale0
\Yboxdim9pt

\section{Introduction}
At the classical level, the thermodynamic properties of ferromagnets and antiferromagnets
are quite similar.  Both states break certain internal symmetries, whether they be
discrete or continuous, and often crystalline point group symmetries as well.
Antiferromagnetism holds the interesting possibility of frustration, which
can lead to complex behavior even at the classical level.

Quantum mechanics further distinguishes antiferromagnetism as the more interesting
of the two phenomena.  Quantum fluctuations compete with classical ordering,
and many models of quantum antiferromagnetism remain disordered even in their
ground states.  The reason is that on the local level, quantum antiferromagnets
prefer distinctly non-classical correlations, in that they form singlets, which are
superpositions of classical states. For a $S=\half$ Heisenberg antiferromagnet
on a bipartite lattice, theorems by Marshall \cite{MAR55} and by Lieb and Mattis
\cite{LM62} rigorously prove that the ground state is a {\it total\/} spin singlet:
$S\nd_{\rm tot}=0$.  Any total singlet can be expanded in a (nonorthogonal) basis
of valence bonds, which are singlet pairs $(ij)\equiv 2^{-1/2}\big(\ket{\!\!\uar\nd_i\,
\dar\nd_j\!}-\ket{\!\!\dar\nd_i\,\uar\nd_j\!}\big)$ extending between sites $i$ and $j$.
The most probable singlets are between nearest neighbors, which thereby take
full advantage of the Heisenberg interaction $J\,\bfS\nd_i\cdot\bfS\nd_j$ and
achieve a minimum possible energy $\ve\nd_0=-\frac{3}{4}\,J$ for that particular link.
Taking linear combinations of such states lowers the energy, via delocalization, with
respect to any fixed singlet configuration; this is the basic idea behind Anderson's
celebrated resonating valence bond (RVB) picture \cite{RVB}.  If one allows the
singlet bonds to be long-ranged, such a state can even possess classical N{\'e}el
order \cite{LDA88}.

Taking advantage of quantum singlets, one can construct correlated quantum-disordered
wavefunctions which are eigenstates of local projection operators.  This feature allows
one to construct a many-body Hamiltonian which renders the parent wavefunction an
exact ground state, typically with a gap to low-energy excitations.  Perhaps the simplest
example is the Majumdar-Ghosh (MG) model for a $S=\half$ spin-chain \cite{MG69}, the
parent state of which is given by alternating singlet bonds, {\it viz.}
\begin{equation}
\ket{\rmPsi}=\ket{\cdots\ \singlet\blank\singlet\blank\singlet\ \cdots}
\end{equation}
The key feature to $\ket{\rmPsi}$ is that any consecutive trio of sites $(n,n+1,n+2)$
can only be in a state of total spin $\cS=\half$ -- there is no $\cS=\frac{3}{2}$ component.
Thus, $\ket{\rmPsi}$ is an eigenstate of the projection operator
\begin{equation}
\rmP\nd_{3/2}(n,n+1,n+2)=-\frac{1}{4}+
\frac{1}{3}\big(\bfS\nd_n+\bfS\nd_{n+1}+\bfS\nd_{n+2} \big)^2
\end{equation}
with zero eigenvalue, and an exact ground state
for $\cH=J\sum_n \rmP\nd_{3/2}(n,n+1,n+2)$.  As $\ket{\rmPsi}$ breaks lattice
translation symmetry, a second (degenerate) ground state follows by shifting
$\ket{\rmPsi}$ by one lattice spacing.  Extensions of the MG model to higher dimensions
and to higher spin, where the ground state is again of the Kekul{\'e} form, \ie\  a product
of local valence bond singlets, were discussed by Klein \cite{KLE82}.

Another example is furnished by the valence bond solid (VBS) states of Affleck,
Kennedy, Lieb, and Tasaki (AKLT) \cite{AKLT}.  The general AKLT state is compactly
written in terms of Schwinger boson operators \cite{AAH88}:
\begin{equation}
\ket{\rmPsi(\cL\,;\,M)}=\prod_{\langle ij\rangle}
\big(b\yd_{i\uar}\,b\yd_{j\dar} - b\yd_{i\dar}\,b\yd_{j\uar}\big)^M\,\ket{0}\ ,
\label{VBS}
\end{equation}
which assigns $m$ singlet creation operators to each link of a lattice $\cL$.  The total
boson occupancy on each site is $zM$, where $z$ is the lattice coordination
number; in the Schwinger representation this corresponds to $2S$.  Thus, a discrete
family of AKLT states with $S=\half z M$ is defined on each lattice, where
$M$ is any integer.  The maximum total spin on any link is then
$S_{ij}^{\rm max}=2S-M$,  and any Hamiltonian constructed out of link projectors
for total spin $\cS\in\big[2S-M+1\, ,\, 2S\big]$, with positive coefficients, renders
$\ket{\rmPsi(\cL\,;\,M)}$ an exact zero energy ground state.  The elementary
excitations in these states were treated using a single mode approximation (SMA) in
ref. \cite{AAH88}.

The ability of two spins to form a singlet state is a special property of the
group SU(2).  Decomposing the product of two spin-$S$ representations yields
the well-known result,
\begin{equation}
S\otimes S=0\oplus 1\oplus 2\oplus\cdots\oplus 2S\ ,
\end{equation}
and there is always a singlet available.  If we replace SU(2) by SU(3), this is no longer
the case.  The representations of SU(N) are classified by $(N-1)$-row Young tableaux
$(l\nd_1,l\nd_2,\, \ldots\, ,l\nd_{N-1})$ with $l\nd_j$ boxes in row $j$, and with $l\nd_1\ge l\nd_2
\ge\cdots\ge l\nd_{N-1}$.  The product of two fundamental $(1,0)$ representations of SU(3) is
\begin{equation}
\mathop{\yng(1)}_{\bf 3}\ \otimes\ \mathop{\yng(1)}_{\bf 3}\ =\ 
\mathop{\yng(1,1)}_{\overline{\bf 3}}\ \oplus\ \mathop{\yng(2)}_{\bf 6}\quad ,
\end{equation}
which does not contain a singlet.  One way to rescue the two-site singlet, for general
SU($N$), is to take the product of the fundamental representation $\bfN$ with the
antifundamental ${\overline \bfN}$.  This yields a singlet plus the $(N^2-1)$-dimensional
adjoint representation.  In this manner, generalizations of the SU(2) antiferromagnet
can be defined in such a manner that the two-site valence bond structure is preserved,
but only on bipartite lattices \cite{AFF85,FOOT}.

Another approach is to keep the same representation of SU($N$) on each site, but
to create SU($N$) singlets extending over multiple sites.  When each site is in the
fundamental representation, one creates $N$-site singlets,
\begin{equation}
\eps^{\alpha\nd_1\ldots\alpha\nd_N}\,b\yd_{\alpha\nd_1}(i\nd_1)\cdots
b\yd_{\alpha\nd_N}(i\nd_N)\,\ket{0}\ ,
\end{equation}
where $b\yd_{\alpha}(i)$ creates a Schwinger boson of flavor index $\alpha$ on site $i$.
The SU($N$) spin operators may be written in terms of the Schwinger bosons as
\begin{equation}
S^\alpha_\beta=b\yd_\alpha\,b\nd_\beta-{p\over N}\,\delta\nd_{\alpha\beta}\ ,
\end{equation}
with ${\rm Tr\,}(S)=0$, for the general symmetric $(p,0)$ representation.
These satisfy the SU($N$) commutation relations,
\begin{equation}
\big[S^\alpha_\beta\, , \, S^\mu_\nu\big]=\delta\nd_{\beta\mu}\,S^\alpha_\nu
-\delta\nd_{\alpha\nu}\,S^\beta_\mu\ .
\end{equation}
Extended valence bond solid (XVBS) states were first discussed by Affleck \etal\ in ref.
\cite{AAMR91}.  In that work, SU($2N$) states where $N=mz$ were defined on
lattices of coordination number $z$, with singlets extending over $z+1$ sites.
Like the MG model, the XVBS states break lattice translation symmetry $t$ and their
ground states are doubly degenerate; they also break a charge conjugation symmetry
$\cC$, preserving the product $t\,\cC$.  In addition to SMA magnons, the XVBS
states were found to exhibit soliton excitations interpolating between the degenerate
vacua.  More recently, Greiter and Rachel \cite{GR07} considered SU($N$) valence bond
spin chains in both the fundamental and other representations, constructing their
corresponding Hamiltonians, and discussing soliton excitations.   Extensions of Klein models,
with Kekul{\'e} ground states consisting of products of local SU($N$) singlets, were discussed
by Shen \cite{SH01}, and more recently by Nussinov and Ortiz \cite{NUS07}.  An SU$(4)$ model on 
a two leg ladder with with a doubly degenerate Majumdar-Ghosh type ground state
has been discussed by Chen {\it et al.\/} \cite{CH05}.

Shen also discussed a generalization of Anderson's RVB state to SU($N$) spins, as a prototype
of a spin-orbit liquid state \cite{SH01}.  A more clearly defined and well-analyzed model was
recently put forward by Pankov, Moessner, and Sondhi \cite{PAN07}, who generalized the
Rokhsar-Kivelson quantum dimer model \cite{ROK88} to a
model of resonating singlet valence plaquettes.  Their plaquettes are $N$-site SU$(N)$ singlets
($N=3$ and $N=4$ models were considered), which resonate under the action of the SU($N$)
antiferromagnetic Heisenberg Hamiltonian, projected to the valence plaquette subspace.
The models and states considered here do not exhibit this phenomenon of resonance.
Rather, they are described by static ``singlet valence simplex" configurations.  Consequently,
their physics is quite different, and in fact simpler.  For example, with resonating valence bonds
or plaquettes, one can introduce vison excitations \cite{SEN01} which are $\rbfZ\nd_2$ vortex
excitations, changing the sign of the bonds or plaquettes which are crossed by the vortex string
\cite{MIS02}.  For simplex (or plaquette) solids, there is no resonance, and the vison does not
create a distinct quantum state.   The absence of `topological quantum order' in Klein and AKLT
models has been addressed by Nussinov and Ortiz \cite{NUS07}.  

Here I shall explore further generalizations of the AKLT scheme, describing a
family of `simplex solid' (SS) states on $N$-partite lattices.  While the general
AKLT state is written as a product over the links of a lattice $\cL$, with $M$
singlet creation operators applied to a given link, the SS states, {\it mutatis mutandis\/},
apply $M$ SU$(N)$ singlet operators on each $N$-simplex.
Each site then contains an SU$(N)$ spin whose representation is determined by $M$
and the lattice coordination.  Furthermore, as is the case with the AKLT states, the SS
states admit a simple coherent state description in terms of classical ${\rm CP}^{N-1}$
vectors.  Their equal-time quantum correlations are then computable as the finite
temperature correlations of an associated classical model on the same lattice.
A classical ordering transition in this model corresponds to a zero-temperature
quantum critical point as a function of $M$ (which is, however, a discrete
parameter).  I argue that the ordered SS states select a particular
ordered structure via an `order by disorder' mechanism.  Finally, I discuss what
happens to these states at an edge, where the bulk SU($N$) representation is effectively
fractionalized, and a residual entropy proportional to the volume of the boundary arises.

\section{Simplex Solids}
Consider an $N$-site simplex $\Gamma$, and define the SU($N$) singlet creation
operator
\begin{equation}
\cR\yd_\Gamma=\eps^{\alpha\nd_1\ldots\alpha\nd_N}\,
b\yd_{\alpha\nd_1}(\Gamma\nd_1)\cdots b\yd_{\alpha\nd_N}(\Gamma\nd_N)\ ,
\end{equation}
where $i=1,\ldots,N$ labels the sites $\Gamma\nd_i$ on the simplex.  Any permutation
$\mu$ of the labels has the trivial consequence of $\cR\yd_\Gamma\to{\rm sgn}(\mu)
\,\cR\yd_\Gamma$.  Next, partition
a lattice $\cL$ into $N$-site simplices, \ie\ into $N$ sublattices, and define the state
\begin{equation}
\ket{\rmPsi(\cL\,;\,M)}=\prod_\Gamma \big(\cR\yd_\Gamma\big)^{\!M}
\,\ket{0}\ ,
\end{equation}
where $M$ is an integer.  Since each $\cR\yd_\Gamma$ operator adds one Schwinger
boson to every site in the simplex, the total boson occupancy of any given site is
$p=\zeta  M$, where $\zeta$ is the number of simplices associated with each site.
For lattices such as the Kagom{\'e} and pyrochlore systems, where two neighboring simplices share a single site, we have $\zeta=2$.  For the tripartite triangular lattice, $\zeta=3$.  Recall that each site is in
the $(p,0)$ representation of SU($N$), with one row of $p$ boxes.

\begin{figure}[!t]
\centering
\includegraphics[width=6.5cm]{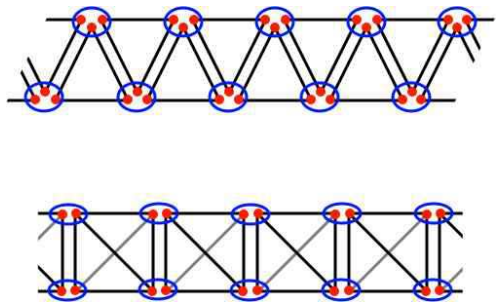}
\caption
{\label{chains} Top: SU(3) SS state on two-leg zigzag chain.  Each site is in the 
10-dimensional totally symmetric representation with 3 boxes (the red dots).
Bottom: SU(4) SS state on two-leg ladder of tetrahedra.
Each site is in the 10-dimensional totally symmetric representation with 2 boxes.}
\end{figure}

\Yboxdim7pt
Two one-dimensional examples are depicted in fig. \ref{chains}.  The first is defined
on a two-leg zigzag chain.  The chain is partitioned into triangles as shown, with each site
being a member of three triangles.  Each triangle represents a simplex $\Gamma$ and
accommodates one power of the SU(3) singlet creation operator $\cR\yd_\Gamma$.
Thus, for this state we have $N=3$ and $\zeta=3$.  With $M=1$ then, the local
SU(3) representation on each site is $(3,0)$, \ie\ $\yng(3)$, which is $10$-dimensional.
The $M=1$ case is in fact a redrawn version of the state defined by Greiter and
Rachel in eqn. 52 of ref. \cite{GR07}.  For this state, a given link may be in any of four
representations of SU(3), or a linear combination thererof:
\Yboxdim9pt
\begin{align}
\mathop{\yng(3)}_{\bf 10}\ \otimes\ \mathop{\yng(3)}_{\bf 10}&\ =\  
\mathop{\yng(3,3)}_{\overline{\bf 10}}\ \oplus\ \mathop{\yng(4,2)}_{\bf 27}\\
&\qquad\ \oplus\ \mathop{\yng(5,1)}_{\bf 35}\ \oplus\ 
\mathop{\yng(6)}_{\bf 28}\ ,\bvph\nonumber
\end{align}
Thus, the zigzag chain SU(3) SS state is an exact zero energy eigenstate of any Hamiltonian of the form
\Yboxdim5pt
\begin{align}
\cH&=\sum_{\langle ij\rangle\atop {\rm sides}} J\nd_1\,\rmP\nd_{\yng(6)}(ij)\\
&\qquad+\sum_{\langle ij\rangle\atop {\rm zigzag}} J\nd_2\, \rmP\nd_{\yng(6)}(ij)
+ J\nd_3\,\rmP\nd_{\yng(5,1)}(ij)\ ,\nonumber
\end{align}
with positive coefficients $J\nd_1$, $J\nd_2$, and $J\nd_3$. 

\Yboxdim7pt
The SU(4) SS chain in fig. \ref{chains} is topologically equivalent to a chain of tetrahedra, 
each joined to the next along an opposite side.  Thus, $\zeta=2$ and for the $M=1$
parent state, each site is in the 10-dimensional \ $\yng(2)$\ \  representation.  From
\Yboxdim9pt
\begin{equation}
\mathop{\yng(2)}_{\bf 10}\ \otimes\ \mathop{\yng(2)}_{\bf 10}\ =
\ \mathop{\yng(2,2)}_{\bf 20}
\ \oplus\ \mathop{\yng(3,1)}_{\bf 45}\ \oplus\ \mathop{\yng(4)}_{\bf 35}\ ,
\label{hundred}
\end{equation}
we can construct a Hamiltonian,
\Yboxdim5pt
\begin{align}
\cH&=\sum_{\langle ij\rangle\atop {\rm sides}} J\nd_1\,\rmP\nd_{\yng(4)}(ij)
+\sum_{\langle ij\rangle\atop {\rm crosses}} J\nd_4\,\rmP\nd_{\yng(4)}(ij)\\
&\qquad+\sum_{\langle ij\rangle\atop {\rm rungs}} J\nd_2\, \rmP\nd_{\yng(4)}(ij)
+ J\nd_3\,\rmP\nd_{\yng(3,1)}(ij)\ ,\nonumber
\end{align}
again with positive coefficients $J\nd_1$, $J\nd_2$, $J\nd_3$ and $J\nd_4$, which renders the
wavefunction $\ket{\rmPsi}$ an exact zero energy ground state.  For both this and the previously
discussed SU(3) chain, the ground state is nondegenerate.

\subsection{SU($N$) Casimirs}
For a collection of $K$ spins, each in the fundamental of SU($N$), we write
\begin{equation}
\cS^\alpha_\beta=\sum_{k=1}^K S^\alpha_\beta(k)\ .
\end{equation}
From the spin operators $\cS^\alpha_\beta$, one can construct $N-1$ Casimirs,
$C^{(n)}={1\over n!}\,{\rm Tr}\,\big(\cS^n\big)$, with $n=2,\ldots,N$.  The eigenvalues of $C^{(n)}$
for totally symmetric ($+$) and totally antisymmetric ($-$) representations of $p$ boxes
were obtained by Kobayashi \cite{KOB73}:
\begin{align}
C^{(n)}(h\,;\,\pm)&={h\,(N\mp1)\,(N\pm p)\over n!\,N^n \,(N\pm p\mp1)}
\cdot\bigg\{(-1)^n\,p^{n-1} +\nonumber\\
&\qquad\qquad  \Big( (N\mp1)\,(N\pm p)\Big)^{n-1}\bigg\}\ .
\end{align}
The Casimirs can be used to construct the projectors onto a given representation as
a polynomial function of the spin operators.  In order to do so, though, we will need
the eigenvalues for all the representations which occur in a given product.
Consider, for example, the case of three SU(3) objects, each in their fundamental
representation.  We then have
\Yboxdim9pt
\begin{equation}
\mathop{\yng(1)}_{\bf 3}\ \otimes\ \mathop{\yng(1)}_{\bf 3}\ \otimes
\ \mathop{\yng(1)}_{\bf 3}\ =\ \mathop{\bullet}_{\bf 1}
\ \oplus\ 2\cdot\mathop{\yng(2,1)}_{\bf 8}\ \oplus\ \mathop{\yng(3)}_{\bf 10}\ .
\end{equation}
The eigenvalues of the quadratic and cubic Casimirs are found to be
\Yboxdim5pt
\begin{align*}
C^{(2)}({\bullet})&=0 \quad& C^{(2)}\big(\,\yng(2,1)\,\big)&=3
\quad& C^{(2)}(\,\yng(3)\,)&=6\vph\\
C^{(3)}({\bullet})&=-4 \quad& C^{(3)}\big(\,\yng(2,1)\,\big)&=0
\quad& C^{(3)}(\,\yng(3)\,)&=8\ .
\end{align*}
Therefore 
\Yboxdim5pt
\begin{align}
\rmP\nd_\bullet(ijk)&=2-\frac{2}{3}\,C^{(2)} + \frac{1}{4}\,C^{(3)}\\
\rmP\nd_{\yng(2,1)}(ijk)&=-2+C^{(2)} -\half\,C^{(3)}\bvph\\
\rmP\nd_{\yng(3)}(ijk)&=1-\frac{1}{3}\,C^{(2)} + \frac{1}{4}\,C^{(3)}\ .
\end{align}
Expressing the projector $\rmP\nd_{\yng(3)}(ijk)$ in terms of the local spin operators
$S^\alpha_\beta(l)$, I find
\begin{align}
\rmP\nd_{\yng(3)}&(ijk)=1-\frac{1}{6}\,{\rm Tr}\,(\cS_{ijk}^2)+ \frac{1}{24}\,{\rm Tr}\,(\cS_{ijk}^3)\label{Pijk}\\
\quad&=-\frac{1}{3}\,{\rm Tr}\,\big[ S(i)\,S(j) + S(j)\,S(k)+ S(k)\,S(i)\big]\nonumber\\
\quad\qquad& +\frac{1}{8}\,{\rm Tr}\,\big[S(i)\,S(j)\,S(k) + S(k)\,S(j)\,S(i)\big]-\frac{2}{27}\ . \nonumber
\end{align}
The projector thus contains both bilinear and trilinear terms in the local spin operators.
One could also write the projector in terms of the quadratic Casimir $C^{(2)}$ only, as
\begin{equation}
\rmP\nd_{\yng(3)}(ijk)=\frac{1}{18}\,C^{(2)}\,\big( C^{(2)}-3\big)\ .
\end{equation}
This, however, would result in interaction terms such as ${\rm Tr}\big[S(i)\,S(j)\big]\cdot
{\rm Tr}\big[S(j)\,S(k)\big]$, which is apparently quadratic in $S(j)$.  For the
\ $\yng(1)$\ representation, however, products such as $S^\alpha_\beta(i)\,S^\mu_\nu(i)$ can
be reduced to linear combinations of the spin operators $S^\sigma_\tau(i)$, as is familiar in the
case of SU(2).  This simplification would then recover the expression in eqn. \ref{Pijk}.

\begin{figure}[!t]
\centering
\includegraphics[width=6.5cm]{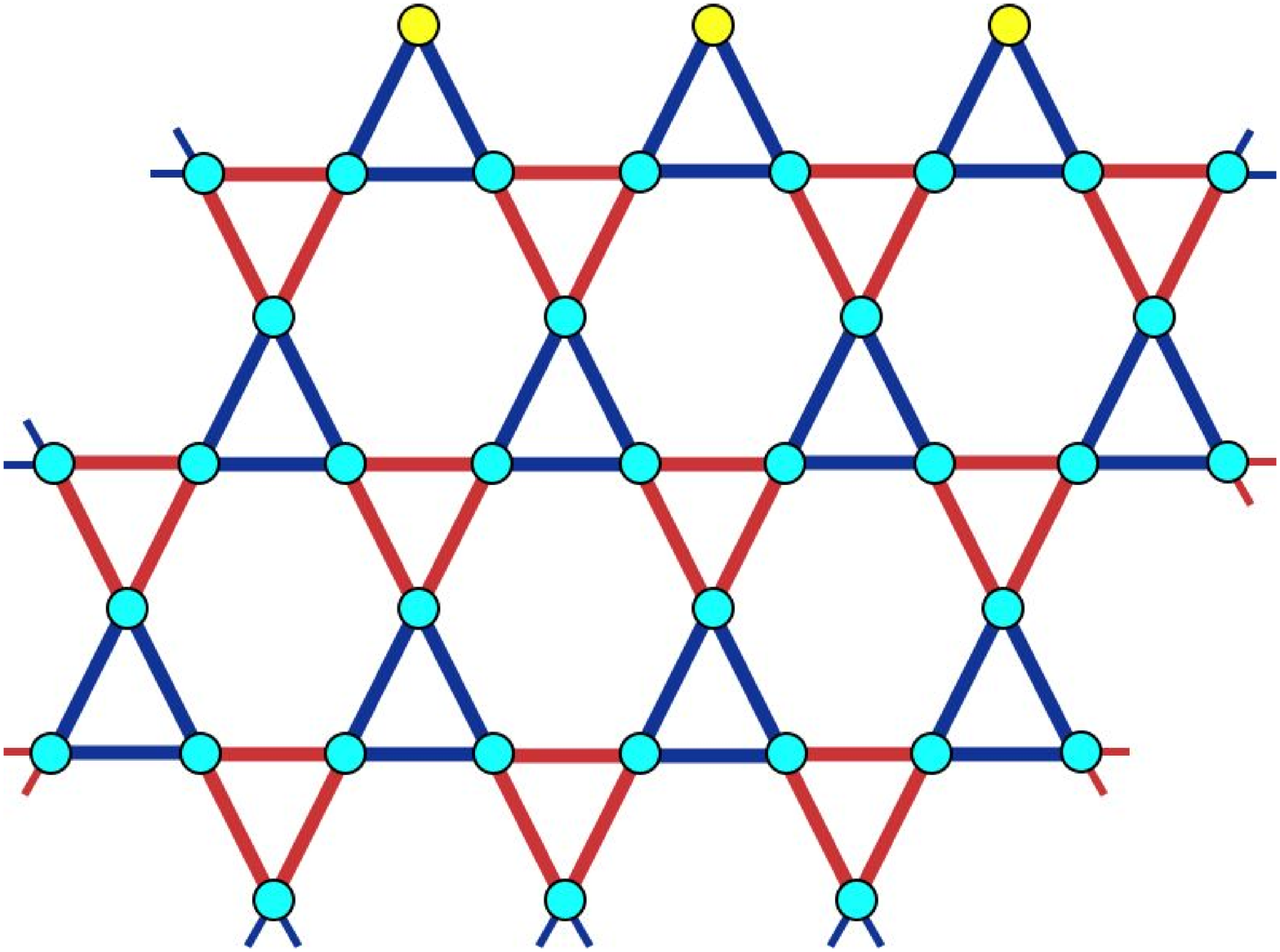}
\caption
{\label{kagome} SU(3) simplex solid states on the Kagom{\'e} lattice.
Applying the singlet operator $\cR\yd_\Gamma$ to all the up (down) triangles
generates the state $\ket{\!\rmPsi_{\!\bigtriangleup\big(\bigtriangledown\big)}\!}$.
Applying $\cR\yd_\Gamma$ to {\it all} the triangular simplices generates the state
$\ket{\rmPsi}$ of eqn. \ref{SSkag}.  The twofold coordinated yellow sites at the top
form a $(10)$ edge (see sec. \ref{edge}).}
\end{figure}

\section{SS states in $d\ge 2$ dimensions}
\subsection{Kagom{\'e} Lattice}
The Kagom{\'e} lattice, depicted in fig. \ref{kagome}, is a two-dimensional network
of corner-sharing triangles, with $\zeta=2$.   It naturally accommodates a set of
$N=3$ SS states.  The simplest example consists of SU(3) objects in the fundamental
representation at each site, and places SU(3) singlets on all the upward-pointing
triangles (see fig. \ref{kagome}):
\begin{equation}
\ket{\rmPsi\nd_\bigtriangleup}=\prod_\bigtriangleup \cR\yd_\bigtriangleup\,\ket{0}\ .
\end{equation}
The lattice inversion operator $\cI$ then generates a degenerate mate,
$\ket{\rmPsi\nd_\bigtriangledown}=\cI\,\ket{\rmPsi\nd_\bigtriangleup}$
Both states are exact zero energy eigenstates of the Hamiltonian
\begin{equation}
\cH=\hskip-8pt\sum_{\langle ijk\rangle\in 120^\circ}\hskip-8pt\rmP\nd_{\yng(3)}(ijk)\ ,
\end{equation}
\Yboxdim7pt
where the sum is over all $120^\circ$ trios $(ijk)$; there are six such $(ijk)$ trios for
every hexagon.  Since two of the three sites in each trio are antisymmetrized,  The fully
symmetric\ \ $\yng(3)$\ \ representation is completely absent.  This model bears
obvious similarities to the MG model: its ground state is a product over independent
local singlets, hence there are no correlations beyond a single simplex, and it
spontaneously breaks a discrete lattice symmetry (in this case $\cI$) \cite{disc}.

If we let the singlet creation operators act on both the up- and down-pointing triangles,
we obtain a state which breaks no discrete lattice symmetries,
\begin{equation}
\ket{\rmPsi}=\prod_\bigtriangleup \cR\yd_\bigtriangleup\,
\prod_\bigtriangledown \cR\yd_\bigtriangledown\,\ket{0}\ .
\label{SSkag}
\end{equation}
For this state, each site is in the 6-dimensional $\yng(2)$ representation.  On any
given link, then, there are the following possibilities:
\Yboxdim9pt
\begin{equation}
\mathop{\yng(2)}_{\bf 6}\ \otimes\ \mathop{\yng(2)}_{\bf 6}
\ =\ \mathop{\yng(2,2)}_{\overline{\bf 6}}
\ \oplus\ \mathop{\yng(3,1)}_{\bf 15}\ \oplus\ \mathop{\yng(4)}_{\bf 15}\ .
\end{equation}
\Yboxdim5pt
The fact that each link belongs to either an $\bigtriangleup$ or $\bigtriangledown$
simplex, and the fact that a singlet operator $\cR\yd_{\bigtriangleup/\bigtriangledown}$
is associated with each simplex, means that no link can be in the fully symmetric
$\yng(4)$ representation.  Thus, $\ket{\rmPsi}$ is an exact, zero-energy eigenstate
of the Hamiltonian
\begin{equation}
\cH=\sum_{\langle ij\rangle}\rmP\nd_{\yng(4)}(ij)\ .
\label{fbox}
\end{equation}
The states $\ket{\rmPsi\nd_\bigtriangleup}$, $\ket{\rmPsi\nd_\bigtriangledown}$, and
$\ket{\rmPsi}$ are depicted in fig. \ref{kagome}. 

The actions of the quadratic and cubic Casimirs on the possible
representations for a given link are given in the following table:
\Yboxdim7pt
\begin{center}
\begin{tabular}{|c||c|c|}
\hline
\hskip2.0in &&\\
Rep$^{\rm n}$ & $C^{(2)}$ & $C^{(3)}$ \\
\phantom{\rm ABCDEFG} & \phantom{\rm ABCDE} & \phantom{\rm ABCDEF} \\
\hline
&&\\
$\yng(2,2)$ & $10/3$ & $-80/27$  \\
&&\\
\hline
&&\\
$\yng(3,1)$ & $16/3$ & $64/27$\\
&&\\
\hline
&&\\
$\yng(4)$ & $28/3$ & $352/27$ \\
&&\\
\hline
\end{tabular}
\end{center}
Note that
\begin{equation}
C^{(3)}=\frac{8}{3}\Big(C^{(2)}-\frac{320}{27}\Big)\ ,
\end{equation}
hence the two Casimirs are not independent here.
We can, however, write the desired projector,
\begin{align}
\rmP\nd_{\yng(4)}&(ij)=\frac{1}{24}\Big(C^{(2)}-\frac{10}{3}\Big)\Big(C^{(2)}-\frac{16}{3}\Big)\\
&=\frac{1}{24} \,\Big({\rm Tr}\,\big[S(i)\,S(j)\big]\Big)^2 + \frac{7}{36}\,{\rm Tr}\,\big[S(i)\,S(j)\big]
+\frac{5}{27}\ ,
\nonumber
\end{align}
as a bilinear plus biquadratic interaction between neighboring spins.
To derive this result, we write $C^{(2)}=\half {\rm Tr}\,\big[S(i)+S(j)\big]^2$, whence
\begin{align}
C^{(2)}&={\rm Tr}\,\big[S(i)\,S(j)\big]+{\rm Tr}\,\big(S^2\big)\nonumber\\
&={\rm Tr}\,\big[S(i)\,S(j)\big]+p\,(N+p-1)-{p^2\over N}\ .
\end{align}

\begin{figure}[!t]
\centering
\includegraphics[width=6.5cm]{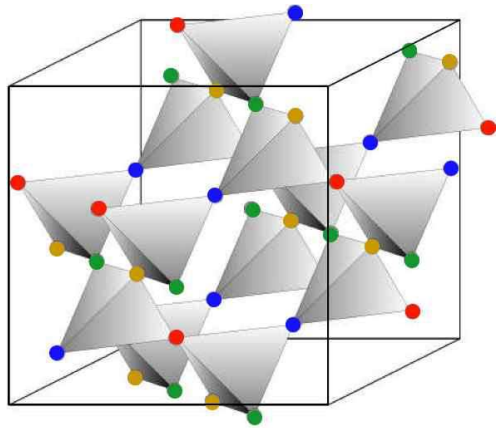}
\caption
{\label{pyrochlore} The pyrochlore lattice and its quadripartite structure.}
\end{figure}

Next, consider the pyrochlore lattice in fig. \ref{pyrochlore}.  This lattice consists
of corner-sharing tetrahedra, with $\zeta=2$, and naturally accommodates an $N=4$
SS state of the form
\begin{equation}
\ket{\rmPsi}=\hskip-7pt \prod_{\rm tetrahedra}\hskip-7pt  \cR\yd_\Gamma\,\ket{0}\ .
\end{equation}
Like the uniform SU($3$) SS state on the Kagom{\'e} lattice, this SU($4$) state 
describes a lattice of spins which are in the $\ \yng(2)\ $ representation on each
site; in the SU($4$) case this representation is 10-dimensional.  From eqn. \ref{hundred}
we see that each link, the sites of which appear in some simplex singlet creation
operator, cannot have any weight in the 35-dimensional totally symmetric
$\ \yng(4)\ $ representation.  Hence, once again, the desired Hamiltonian is
that of eqn. \ref{fbox}.  For SU(4), 

\Yboxdim5pt
\begin{equation*}
C^{(2)}\big(\,\yng(2,2)\,\big)=6 \quad,\quad C^{(2)}\big(\,\yng(3,1)\,\big)=8
\quad,\quad C^{(2)}\big(\,\yng(4)\,\big)=12\ ,
\end{equation*}
and so
\begin{align}
\rmP\nd_{\yng(4)}&(ij)=\frac{1}{24}\big(C^{(2)}-6\big)\big(C^{(2)}-8\big)\\
&=\frac{1}{24} \,\Big({\rm Tr}\,\big[S(i)\,S(j)\big]\Big)^2 + \frac{1}{6}\,{\rm Tr}\,\big[S(i)\,S(j)\big]
+\frac{1}{8}\ ,
\nonumber
\end{align}

Indeed, there is a rather direct correspondence between the possible SU($3$)
SS states on the Kagom{\'e} lattice and the SU($4$) SS states on the pyrochlore
lattice.  For example, one can construct a model with a doubly degenerate
ground state, similar to the MG model, by associating the simplex singlet
operators $\cR\yd_\Gamma$ with only the tetrahedra which point along the
$(111)$ lattice direction.

\begin{figure}[!t]
\centering
\includegraphics[width=6cm]{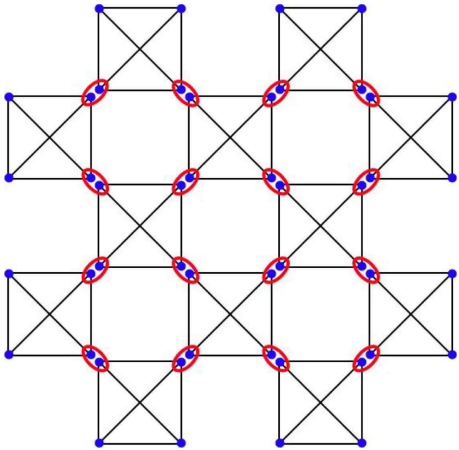}
\caption
{\label{su4sq} One of two doubly degenerate ground states for the Hamiltonian of eqn. 
\ref{fbox} applied to the square lattice, where each site is in the $(2,0,0)$ representation 
of SU(4).  The squares with crosses (or, equivalently, tetrahedra) represent singlet operators
$\eps^{\alpha\beta\gamma\delta}\,b\yd_\alpha(i)\,b\yd_\beta(j)\,b\yd_\gamma(k)\,b\yd_\delta(l)$
on the simplex $(ijkl)$.  The resulting `planar pyrochlore' configuration is equivalent to a checkerboard
lattice of tetrahedra.}
\end{figure}

Finally, consider SU(4) states on the square lattice, again in the \ $\yng(2)$ \ representation
on each site.  We can once again identify the exact ground state of the $\rmP\nd_{\yng(4)}(ij)$
Hamiltonian of eqn. \ref{fbox}.  In this case, the ground state is doubly degenerate, and is
described by the `planar pyrochlore' configuration shown in fig. \ref{su4sq}.

\section{Mapping to a Classical Model}
The correlations in the SS states are calculable using the coherent state representation.
From results derived in the Appendix I,  the coherent state SS wavefunction is given by
\begin{equation}
\rmPsi\big[\{\zbar(i)\}\big]=\cC\>\prod_\Gamma \left[R\nd_\Gamma
\big(\zbar(\Gamma\nd_1)\, ,\, \ldots\, ,\, \zbar(\Gamma\nd_N)\big)\right]^M\ ,
\end{equation}
where $\cC$ is a normalization constant, and
\begin{align}
R\nd_\Gamma&\equiv \zbar(\Gamma\nd_1)\wedge\zbar(\Gamma\nd_2)\wedge
\cdots\wedge\zbar(\Gamma\nd_N)\nonumber\\
&=\eps^{\alpha\nd_1\ldots\alpha\nd_N}\, \zbar_{\alpha\nd_1}(\Gamma\nd_1)\cdots
\zbar_{\alpha\nd_N}(\Gamma\nd_N)\ .
\end{align}
Here I have labeled the $N$ sites on each simplex $\Gamma$ by an index
$i$ running from $1$ to $N$.

Note that the coherent state probability density is
\begin{equation}
\big| \rmPsi\big|^2 = |\cC|^2\,\prod_\Gamma \big| R\nd_\Gamma\big|^{2M}\ ,
\end{equation}
and that
\begin{align}
\big| R\nd_\Gamma\big|^2=\eps^{\alpha\nd_1\cdots\alpha\nd_N}\,
\eps^{\beta\nd_1\cdots\beta\nd_N}\,Q\nd_{\alpha\nd_1\beta\nd_1}\!(\Gamma\nd_1)
\cdots Q\nd_{\alpha\nd_N\beta\nd_N}\!(\Gamma\nd_N)
\end{align}
where $Q\nd_{\alpha\beta}(i)=\zbar\nd_\alpha(i)\,z\nd_\beta(i)$.
Writing $|\rmPsi|^2\equiv e^{-\Hcl/T}$, we see that that the probability
density may be written as the classical Boltzmann weight for a system described by
the classical Hamiltonian
\begin{equation}
\Hcl=-\sum_\Gamma\ln \big|R\nd_\Gamma\big|^2\ ,
\label{Ham}
\end{equation}
at a temperature $T=1/M$ \cite{LAU83,AAH88}.  The classical interactions are
$N$-body interaction, involving the matrices $Q\nd_{\alpha\beta}(i)$ on all the sites
of a given $N$-site simplex, summed over all distinct simplices.  For $N=2$, this
results in a classical nearest-neighbor quantum antiferromagnet \cite{AAH88}, with
\begin{equation}
H^\ssr{AKLT}_\ssr{cl}=-\sum_{\langle ij\rangle} \ln\bigg({1-\nhat\nd_i\cdot\nhat\nd_j\over 2}\bigg)\ ,
\label{AKLTcl}
\end{equation}
with $\nhat=z\yd\bfsigma z$, where $\bfsigma$ is the vector of Pauli matrices.
This general feature of pair product wavefunctions of the Bijl-Feynman, Laughlin, and
AKLT form is thus valid for the SS states as well.

As shown in Appendix I, the matrix element $\expect{\phi}{{\hat T}\nd_K}{\psi}$
of an operator
\begin{equation}
{\hat T}\nd_K={\sum_{\bfm,\bfn}}'\ T\nd_{\bfk,\bfl}\ {b\nd_1}^{\!\!\!k\nd_1}\cdots
{b\nd_N}^{\!\!\!\!\!k\nd_N}\>{b\yd_1}^{l\nd_1}\cdots{b\yd_N}^{\!\!l\nd_N}
\label{TKO}
\end{equation}
may be computed as an integral with respect to the measure $d\mu$ (on each site)
of the product ${\bar\phi}(z)\,\psi(\zbar)$ of coherent state wavefunctions multiplied
by the kernel
\begin{equation}
{\hat T}\nd_K\big(\{b\nd_\alpha\},\{b\yd_\alpha\}\big)\rightarrow
{\big(N-1+p+K\big)!\over p\,!}\ {\hat T}\nd_K\big(\{z\nd_\alpha\},\{\zbar\nd_\alpha\}\big)\ .
\end{equation}
Thus, the {\it quantum mechanical\/} expectation values of Hermitian observables in the
SS states are expressible as {\it thermal\/} averages over the corresponding classical
Hamiltonian $\Hcl$ of eqn. \ref{Ham}.  The SS and VBS states thus share the special
property that their equal time quantum correlations are equivalent to thermal
correlations of an associated classical model on the same lattice, \ie\ in the same
number of dimensions.

In this paper I will be content to merely elucidate the correspondence between
quantum correlations in $\ket{\rmPsi(\cL\,;\,M})$ and classical correlations in
$\Hcl$.  An application of this correspondence to a Monte Carlo evaluation of the
classical correlations will be deferred to a future publication.

\section{Single Mode Approximation for Adjoint Excitations}
Following the treatment in ref. \cite{AAH88}, I construct trial excited states at wavevector $\bfk$
in the following manner.  First, define the operator
\begin{align}
\phi\nd_{\alpha\beta}(\bfk)&=\sum_i \eta\nd_i \,\phi\nd_{i,\alpha\beta}(\bfk)\\
\phi\nd_{i,\alpha\beta}(\bfk)&=\cN^{-1/2}\sum_{\bfR} e^{i\bfk\cdot\bfR}\,S^\alpha_\beta(\bfR,i)\ ,
\end{align}
where $\bfR$ is a Bravais lattice site and $i$ labels the basis elements.  Here, $\eta\nd_i$ is
for the moment an arbitrary set of complex-valued parameters and $\cN$ is the total number
of unit cells in the lattice.  The operators $\phi\nd_{\alpha\beta}(\bfk)$ transform according to the
$(N^2-1)$-dimensional adjoint representation of SU($N$).  Next, construct the trial state,
\begin{equation}
\ket{\! \phi \!}\equiv \phi\nd_{\alpha\beta}(\bfk)\,\ket{\!\rmPsi\!}\ .
\end{equation}
and evaluate the expectation value of the Hamiltonian in this state:
\begin{equation}
E\nd_\ssr{SMA}(\bfk)={\expect{ \! \phi \! } {\!H\!} { \! \phi \!} \over \braket { \! \phi \! } { \! \phi \! } }
={\eta^*_i\, f\nd_{ij}(\bfk)\,\eta\nd_j \over \eta^*_i\,s\nd_{ij}(\bfk)\,\eta\nd_j}\ .
\end{equation}
Here, $f\nd_{ij}(\bfk)$ and $s\nd_{ij}(\bfk)$ are, respectively, the oscillator strength and
structure factor, given by
\begin{align}
f\nd_{ij}(\bfk)&=\half\expect{\!\rmPsi\!} {\Big[\phi\yd_{i,\alpha\beta}(\bfk)\, ,
\big[H\, , \, \phi\nd_{j,\alpha\beta}(\bfk)\big] \Big]} {\! \rmPsi\!}\\
s\nd_{ij}(\bfk)&=\expect{\!\rmPsi\!}{\phi\yd_{i,\alpha\beta}(\bfk)\,\phi\nd_{j,\alpha\beta}(\bfk)}{\!\rmPsi\!}\ .
\end{align}
Here I have assumed that $H$ is a sum of local projectors, and that $H\,\ket{\!\rmPsi\!}=0$.
Treating the $\eta\nd_i$ parameters variationally, one obtains the equation
\begin{equation}
f\nd_{ij}(\bfk)\,\eta\nd_j=E\nd_\ssr{SMA}(\bfk)\,s\nd_{ij}(\bfk)\,\eta\nd_j\ .
\end{equation}
The lowest eigenvalue of this equation provides a rigorous upper bound to the lowest excitation
energy at wavevector $\bfk$.  The result is exact if all the oscillator strength is saturated by a 
single mode, whence the SMA label.  When $\ket{\!\rmPsi\!}$ is quantum-disordered, the SMA 
spectrum is gapped.  When $\ket{\!\rmPsi\!}$ develops long-ranged order (for sufficiently large
$M$ parameter, and in $d>2$ dimensions), the SMA spectrum is gapless.

\section{Mean field treatment of quantum phase transition}
The classical Hamiltonian of eqn \ref{Ham} exhibits a global SU($N$) symmetry, 
where $z\nd_\sigma(i)\to U\nd_{\sigma\sigma'}\,z\nd_{\sigma'}(i)$ for every
lattice site $i$.  Since the interactions are short-ranged, there can be no
spontaneous breaking of this symmetry in dimensions $d\le 2$.  In higher
dimensions, the classical model can order at finite temperature, corresponding
to a quantum ordering at a finite value of $m$.  For the AKLT states, where $N=2$,
this phase transition was first discussed in ref. \cite{AAH88}.  I first discuss the
$N=2$ case and then generalize to arbitrary $N>2$.

\subsection{$N=2$ : VBS States}
Consider the $N=2$ case, which on a lattice of coordination number $z$
yields a family of wavefunctions describing $S=\half M z$ objects with
antiferromagnetic correlations.  These are the AKLT valence bond solid (VBS)
states.  We have
\begin{equation}
z=\begin{pmatrix} \cos\!\big({\theta\over 2}\big) \\ \\ \sin\!\big({\theta\over 2}\big)
\,e^{i\phi}\end{pmatrix}\quad,\quad
Q={1\over 2}\begin{pmatrix} 1+n^z & n^+ \\ & \\ n^- & 1-n^z
\end{pmatrix}\ ,
\end{equation}
where $\bfn=z\yd\bfsigma z$ is a real unit vector, ($\bfsigma$ are the
Pauli matrices).  Since
\begin{equation}
\eps^{\alpha\nd_1\beta\nd_1}\,\eps^{\alpha\nd_2\beta\nd_2}\,
Q\nd_{\alpha\nd_1\beta\nd_1}\!(\Gamma\nd_\ssr{A})\,
Q\nd_{\alpha\nd_2\beta\nd_2}\!(\Gamma\nd_\ssr{B})
=\half\big(1-\nhat\nd_\ssr{A}\cdot\nhat\nd_\ssr{B}\big)\ ,
\end{equation}
the effective Hamiltonian is
\begin{equation}
\Hcl=-\sum_{\langle ij\rangle}\ln\bigg({1-\nhat\nd_i\cdot\nhat\nd_j\over 2}\bigg)\ .
\end{equation}
The sum is over all links on the lattice.  I assume the lattice is bipartite, so each link
connects sites on the A and B sublattices.  I now make the mean field {\it Ansatz\/} 
\begin{equation}
\nhat\nd_\ssr{A}= \bfm+\delta\nhat\nd_\ssr{A}\quad,\quad
\nhat\nd_\ssr{B}= -\bfm+\delta\nhat\nd_\ssr{B}
\end{equation}
and expand $H$ in powers of $\delta\nhat\nd_i$.  Expanding to lowest nontrivial
order in the fluctuations $\delta\nhat\nd_i$, we obtain a mean field Hamiltonian
\begin{equation}
H^\ssr{MF}=E\nd_0-\bfh\nd_\ssr{A}\sum_{i\in \rmA}\nhat\nd_i
-\bfh\nd_\ssr{B}\sum_{j\in \rmB}\nhat\nd_j\ ,
\end{equation}
where $E\nd_0$ is a constant and 
\begin{equation}
\bfh\nd_\ssr{A}=-\bfh\nd_\ssr{B}={z\,\bfm\over 1+\bfm^2}
\end{equation}
is the mean field.  Here $z$ is the lattice coordination number.  The self-consistency
equation is then
\begin{equation}
\bfm=\langle \nhat\nd_\ssr{A}\rangle=\int\!\!d\nhat\>\nhat\>e^{\bfh\nd_\rmA\cdot
\nhat/T}\!\Bigg/\!\!\int\!\!d\nhat\>e^{\bfh\nd_\rmA\cdot\nhat/T}\ ,
\end{equation}
which yields
\begin{equation}
m=\ctnh\lambda - {1\over\lambda}\quad;\quad\lambda={z\over T}\,{m\over 1+m^2}\ .
\end{equation}
The classical transition occurs at $T^\ssr{MF}_\rmc=\frac{1}{3}z$, so the SS state
exhibits a quantum phase transition at $M^\ssr{MF}_\rmc=3 z^{-1}$.  For
$M>M\nd_\rmc$ the SS exhibits long-ranged two-sublattice N{\'e}el order. 
On the cubic lattice, the mean field value $M^\ssr{MF}_\rmc=\frac{1}{2}$ suggests
that all the square lattice VBS states, for which the minimal spin is $S=3$ (with
$M=1$), are N{\'e}el ordered.  Since the mean field treatment overestimates
$T\nd_\rmc$ due to its neglect of fluctuations, I conclude that the true
$M\nd_\rmc$ is somewhat greater than $3z^{-1}$, which leaves open the
possibility that the minimal $M=1$ VBS state on the cubic lattice is a quantum
disordered state.  Whether this is in fact the case could be addressed by a classical
Monte Carlo simulation.

\subsection{$N>2$ : SS States}
For general $N$, I write
\begin{equation}
Q\nd_{\alpha\beta}(i)=\langle\, Q\nd_{\alpha\beta}(i)\,\rangle + \delta
Q\nd_{\alpha\beta}(i)\ ,
\end{equation}
where the average is taken with respect to $|\rmPsi|^2 = e^{-\Hcl/T}$.  To maximize
$|\rmPsi|^2$, \ie\ to minimize $\Hcl$, choose a set
$\big\{\rmP^\sigma_{\alpha\beta}\big\}$ of $N$ mutually orthogonal projectors, with
$\sigma=1,\ldots,N$.  The projectors satisfy the relations
\begin{equation}
\rmP^\sigma\,\rmP^{\sigma'}=\delta\nd_{\sigma\sigma'}\,\rmP^\sigma\ ,
\end{equation}
and can each be written as
\begin{equation}
\rmP^\sigma_{\alpha\beta}={\bar \omega}^\sigma_\alpha\,\omega^\sigma_\beta\ ,
\end{equation}
where $\big\{\omega^\sigma\big\}$ is a set of $N$ mutually orthogonal
$\rmC\rmP^{N-1}$ vectors.
Then if $z(\Gamma\nd_\sigma)=\omega^\sigma$ for each site $\Gamma\nd_\sigma$
in the simplex, we have $R\nd_\Gamma=e^{i\eta}$, where
$\eta$ is an arbitrary phase, and $|R\nd_\Gamma|^2=1$.  One then writes
\begin{equation}
\cQ^\sigma_{\alpha\beta}\equiv\langle\, Q\nd_{\alpha\beta}(\Gamma\nd_\sigma)\,\rangle=
{1\over N}\>\delta\nd_{\alpha\beta}
+ m\,\Big( \rmP^\sigma_{\alpha\beta}-{1\over N}\>\delta\nd_{\alpha\beta}\Big)\ .
\label{Qeqn}
\end{equation}
Here $m\in [0,1]$ is the order parameter, analogous to the magnetization.
When $m=0$, no special subspace is selected, and the correlations are isotropic.
When $m=1$, the $Q$-matrix is a projector onto the one-dimensional subspace
defined by $\omega^\sigma$.  Note that $\langle\,{\rm Tr}\,Q(\Gamma\nd_\sigma)\,\rangle=1$, as it must be.

At this point, there remains a freedom in assigning the vectors $\{\omega^\sigma\}$ to
the  sites $\{\Gamma\nd_\sigma\}$ of each simplex.  Consider, for example, the $N=3$
case on the Kagom{\'e} or triangular lattice.  The lattice is tripartite, so every A sublattice
site has 2 (Kagom{\'e}) or 3 (triangular) nearest neighbors on each of the B and C
sublattices.  However, as is well-known, the individual sublattices may have lower
translational symmetry than the underlying triangular Bravais lattice.  Indeed, the
sublattices may be translationally disordered.  I shall return to this point below.
For the moment it is convenient to think in terms of $N$ sublattices each of which
has the same discrete symmetries as the underlying Bravais lattice.

Expanding $\Hcl$ to lowest order in the fluctuations $\delta Q\nd_{\alpha\beta}(i)$
on each site, and dropping terms of order $(\delta Q)^2$ and higher, I obtain the
mean field Hamiltonian,
\begin{equation}
H_\ssr{MF}=E\nd_0-\zeta\>\sum_i h\nd_{\alpha\beta}(i)\,Q\nd_{\alpha\beta}(i)\ ,
\end{equation}
where $E\nd_0$ is a constant.    The mean field $h\nd_{\alpha\beta}(i)$ is site-dependent.
On a $\Gamma\nd_N$ site, we have
\begin{equation}
h^{(N)}_{\alpha\nd_N\beta\nd_N}={\eps^{\alpha\nd_1\alpha\nd_2
\ldots\alpha\nd_N}\,\eps^{\beta\nd_1\beta\nd_2\ldots\beta\nd_N}\,
\cQ^{(1)}_{\alpha\nd_1\beta\nd_1}\!\!\cdots\cQ^{(N-1)}_{\alpha\nd_{N-1}\beta\nd_{N-1}}
\over \eps^{\mu\nd_1\ldots\mu\nd_N}\,\eps^{\nu\nd_1\ldots\nu\nd_N}\,
\cQ^{(1)}_{\mu\nd_1\nu\nd_1}\!\!\cdots\cQ^{(N)}_{\mu\nd_N\nu\nd_N}}\ ,
\label{lmf}
\end{equation}
In Appendix II I show that
\begin{equation}
h^\sigma_{\alpha\beta}=\Big(A\nd_N(m)\,\delta\nd_{\alpha\beta}
+ B\nd_N(m)\,\rmP^\sigma_{\beta\alpha}\Big)\Big/R\nd_N(m)\ ,
\label{heqn}
\end{equation}
where
\begin{align}
A\nd_N(m)&=(N-2)\,!\>\sum_{j=0}^{N-2}{N-j-1\over j\,!}\ m^j\,
\bigg({1-m\over N}\bigg)^{\!\!N-j-1}\label{Aeqn}\\
B\nd_N(m)&=(N-2)\,!\>\sum_{j=0}^{N-2}{m^{j+1}\over j\,!}\,
\bigg({1-m\over N}\bigg)^{\!\!N-j-2}\bvph\label{Beqn}\\
R\nd_N(m)&=N!\>\sum_{j=0}^N {m^j\over j!}\,\bigg({1-m\over N}\bigg)^{\!\!N-j}\ .
\label{Reqn}
\end{align}
Note that $B\nd_N(0)=0$, $B\nd_N(1)=1$, and $R(1)=1$.

The mean field Hamiltonian is then
\begin{align}
H\nd_\ssr{MF}&=-\sum_i {\rm Tr}\,\Big( h^\rmt(i)\,Q(i)\Big)\nonumber\\
&=-\zeta\,B\nd_N(m)\sum_i \big| \omega\yd(i)\,z(i)\big|^2\ ,
\end{align}
where $h\nd_{\alpha\beta}(i)=h_{\alpha\beta}^{\sigma(i)}$, where 
$\sigma(i)$ labels the projector associated with site $i$.
The self-consistency relation is obtained by evaluating the thermal average
of $Q\nd_{\alpha\beta}(i)$.  With $x\nd_\alpha\equiv |z\nd_\alpha|^2$, I obtain
\begin{align}
{1\over N} + {N-1\over N}\>m &={\int\limits_0^1\!dx\>x\,(1-x)^{N-2}
\exp(\zeta\, b\nd_N(m)\, x/T)\over\int\limits_0^1\!dx\>(1-x)^{N-2}\,
\exp(\zeta\, b\nd_N(m)\, x/T)}\ ,
\end{align}
where $b\nd_N(m)=B\nd_N(m)/R\nd_N(m)$.
It is simple to see that $m=0$ is a solution to this mean field equation.  To find
the critical temperature $T\nd_\rmc$,  expand the right hand side in powers
of $m$ for small $m$.  To lowest order, one finds
\begin{equation}
b\nd_N(m)={N\over N-1}\>m + \cO(m^2)\ .
\end{equation}
The value of $T^\ssr{MF}_\rmc$ is determined by equating the coefficients of $m$ on
either side of the equation.  I find
\begin{equation}
M^\ssr{MF}_\rmc(N,\zeta)={1\over T^\ssr{MF}_\rmc(N,\zeta)}={N^2-1\over\zeta}\ .
\end{equation}
This agrees with our previous result $M_\rmc^\ssr{MF}=\frac{1}{2}$ for the $N=2$
state on the cubic lattice, for which $\zeta=6$.
The $N=3$ SS on the Kagom{\'e} lattice cannot develop long-ranged order
which spontaneously breaks SU($N$), owing to the Mermin-Wagner theorem.
For the $N=4$ SS on the pyrochlore lattice, our mean field theory analysis
suggests that the SS states are quantum disordered up to $M\approx \frac{15}{2}$.
Note that expression for $M^\ssr{MF}_\rmc$ reflects a competition between
fluctuation effects, which favor disorder, and the coordination number, which favors
order.  The numerator, $N^2-1$, is essentially the number of directions
in which $Q$ can fluctuate about its average $\cQ$; this is the dimension of
the Lie algebra su($N$).

\section{Order by Disorder}
At zero temperature, the classical model of eqn. \ref{Ham} is solved by maximizing
$\big|R\nd_\Gamma\big|^2$ for each simplex $\Gamma$.  This is accomplished
by partitioning the lattice $\cL$ into $N$ sublattices, such that no neighboring sites
are elements of the same sublattice.  One then chooses
any set of $N$ mutually orthogonal vectors $\omega\nd_\sigma\in
{\rm CP}^{N-1}$, and set $z\nd_i=\omega\nd_{\sigma(i)}$.  On every $N$-site
simplex, then, each of the $\omega\nd_\sigma$ vectors will occur exactly once,
resulting in $\big|R\nd_\Gamma\big|^2=1$, which is the largest possible value.
Thus, the model is unfrustrated, in the sense that every simplex $\Gamma$ is fully
satisfied by the $z\nd_i$ assignments, and the energy is the minimum possible
value: $E\nd_0=0$.

For $N=2$, there are two equivalent ways of partitioning a bipartite lattice into two
sublattices.  For $N>2$, there are in general an infinite number of inequivalent partitions,
all of which have the same ground state energy $E\nd_0=0$.  At finite temperature, 
though, the free energy of these different orderings will in general differ, due to the
differences in their respective excitation spectra.  A particular partition may then
be selected by entropic effects.  This phenomenon is known as `order by disorder'
\cite{VIL80,SHE82,HEN89}.

\begin{figure}[!t]
\centering
\includegraphics[width=6.5cm]{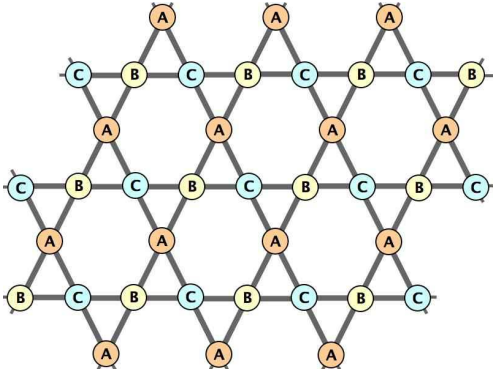}
\caption
{\label{kag_Q0} The $\bfQ=0$ structure on the Kagom{\'e} lattice.}
\end{figure}

To see how entropic effects might select a particular partitioning, I derive
a nonlinear $\sigma$-model, by expanding $\Hcl$ about a particular zero-temperature
ordered state.  Start with
\begin{equation}
z(i)=\omega\nd_{\sigma(i)}\,\big(1-\pi\yd_i\,\pi\nd_i\big)^{1/2}+\pi\nd_i\ ,
\end{equation}
where $\pi\yd_i\, \omega\nd_{\sigma(i)}=0$.  We may now expand
\begin{align}
|R\nd_\Gamma|^2&=\big|\eps^{\alpha\nd_1\cdots\alpha\nd_N}\,z_{\alpha\nd_1}
(\Gamma\nd_1)\cdots z_{\alpha\nd_N}(\Gamma\nd_N)\big|^2\nonumber\\
&=1-\half\sum_{i,j}\big|\,\pi\yd_i \,\omega\nd_{\sigma(j)} + \omega\yd_{\sigma(i)}\,\pi\nd_j
\big|^2 + \ldots
\end{align}
Thus, the `low temperature' classical Hamiltonian is
\begin{equation}
\HLT=\sum_{\langle ij\rangle} \big|\,\pi\yd_i \,\omega\nd_{\sigma(j)} +
\omega\yd_{\sigma(i)}\,\pi\nd_j\big|^2 + \ldots\ ,
\label{HLT}
\end{equation}
where the sum is over all nearest neighbor pairs on the lattice.
The full SU$(N)$ symmetry of the model is of course not apparent in eqn. \ref{HLT},
since it is realized nonlinearly on the $\pi\nd_i$ vectors.

\begin{figure}[!t]
\centering
\includegraphics[width=6.5cm]{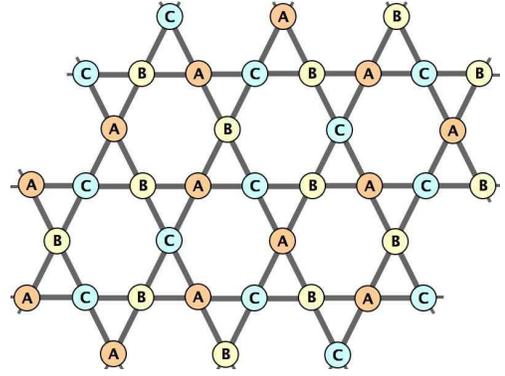}
\caption
{\label{kag_rt3} The $\sqrt{3}\times\sqrt{3}$ structure on the Kagom{\'e} lattice.}
\end{figure}

Each $\pi\nd_i$ vector is subject to a nonholonomic constraint, $\pi\yd_i\,\pi\nd_i
\le 1$.  To solve for the thermodynamics of $\HLT$, I will adopt a simplifying
approximation, in which there is just one nonholonomic constraint,
$\sum_i \pi\yd_i\,\pi\nd_i \le \cN$, where $\cN$ is the number of sites in the lattice.
I fix the constraint by introducing an auxiliary variable $\xhi$ and demanding
\begin{equation}
\cN\,|\xhi|^2+ \sum_i \pi\yd_i\,\pi\nd_i= \cN\ ,
\end{equation}
which I enforce with a Lagrange multiplier $\lambda$.  The resulting model is
\begin{equation}
\HLT=\sum_{\langle ij\rangle} \big|\,\pi\yd_i \,\omega\nd_{\sigma(j)} +
\omega\yd_{\sigma(i)}\,\pi\nd_j\big|^2 + \lambda\,\Big( \cN \, |\xhi|^2 + 
\sum_i \pi\yd_i\,\pi\nd_i-\cN\Big)\ .
\end{equation}
The local constraints $\pi\yd_i\, \omega\nd_{\sigma(i)}=0$ are retained.

It is convenient to rotate to a basis where $\omega^\sigma_\alpha=
\delta\nd_{\alpha,\sigma}$, in which case $\omega\yd_{\sigma(i)}\,\pi\nd_j=
\pi\nd_{j,\sigma(i)}$, \ie\ the $\sigma(i)$ component of the vector $\pi\nd_i$.
So long as $\sigma(i)\ne\sigma(j)$ for nearest neighbors $i$ and $j$, the local
constraints have no effect on the Hamiltonian.  We are then left with a Gaussian
theory in the $\pi\nd_i$ vectors.  Leaving the constraint term aside for the moment,
we can solve for the spectrum of the first part of $\HLT$.  From this spectrum, we
compute the density of states per site, $g(\ve)$.  The free energy per site is then
\begin{equation}
{F\over\cN}=-\lambda + \lambda\,|\xhi|^2 + T\!\int\limits_0^\infty\!\!d\ve\,g(\ve)
\,\ln\!\bigg({\ve+\lambda\over T}\bigg)\ .
\end{equation}
Since $\cN$ is thermodynamically large, we can extremize with respect to $\lambda$,
to find the saddle point, yielding
\begin{equation}
1=|\xhi|^2 + T\!\int\limits_0^\infty\!\! d\ve\,{g(\ve)\over\ve+\lambda}\ .
\end{equation}
Setting $\lambda=\xhi=0$, I obtain an equation for $T\nd_\rmc$,
\begin{equation}
T\nd_\rmc=\Bigg[\int\limits_0^\infty\!\!d\ve\ {g(\ve)\over\ve}\Bigg]^{-1}\ .
\end{equation}
For $T<T\nd_\rmc$, there is Bose condensation, and $|\xhi|^2 > 0$.
For $T>T\nd_\rmc$, the system is disordered.  I stress that the disordered phase,
described in this way, does not reflect the SU$(N)$ symmetry which must be present,
owing to the truncation in eqn. \ref{HLT}.

A natural setting to investigate the order by disorder mechanism would be the SU(4)
SS model on the pyrochlore lattice.  I defer this analysis, together with a companion
Monte Carlo simulation, to a later publication.  Here I will provide a simpler
analysis of the SU(3) Kagom{\'e} SS.  Since the Mermin-Wagner theorem precludes
spontaneous breaking of SU(3) in two dimensions, this analysis at best will reveal
which correlations should dominate at the {\it local} level.  The two structures I
wish to compare are the $\bfQ=0$ structure, depicted in fig. \ref{kag_Q0}, and the
$\sqrt{3}\times\sqrt{3}$ structure, depicted in fig. \ref{kag_rt3}.  Here, the A, B, and C
sites correspond to ${\rm CP}^2$ vectors,
\begin{equation}
\omega\nd_\ssr{A}=\begin{pmatrix} 1 \\ 0 \\ 0 \end{pmatrix} \quad,\quad
\omega\nd_\ssr{B}=\begin{pmatrix} 0 \\ 1 \\ 0 \end{pmatrix} \quad,\quad
\omega\nd_\ssr{C}=\begin{pmatrix} 0 \\ 0 \\ 1 \end{pmatrix}\ .
\end{equation}
Both the $\bfQ=0$ and $\sqrt{3}\times\sqrt{3}$ structures are unfrustrated, in the 
sense that the interactions are fully satisfied on every simplex -- $|R\nd_\Gamma|^2=1$
for all $\Gamma$.  Entropic effects, however, should favor one of the two configurations.

\begin{figure}[!t]
\centering
\includegraphics[width=6.5cm]{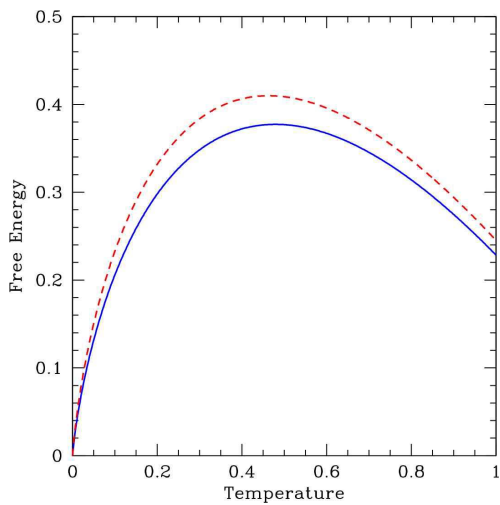}
\caption
{\label{ferg} Free energy for SU(3) simplex solid states on the Kagom{\'e} lattice.
Dashed (red): $\bfQ=0$ structure.  Solid (blue): $\sqrt{3}\times \sqrt{3}$ structure.}
\end{figure}

The $\bfQ=0$ structure may be regarded as a triangular Bravais lattice with a three
element basis (\eg\ a triangular lattice of up-triangles).
Each triangular simplex contains three $\pi$ vectors, each of which
has two independent components (neglecting for the moment the global constraint).
Solving for the spectrum in the absence of the constraint, one finds six branches:
\begin{equation}
\ve\nd_{l,\pm}(\bfk)=2\pm2\cos\big(\half\bfk\cdot\bfa\nd_l\big)\ ,
\end{equation}
where $\bfk$ is a vector in the Brillouin zone, and the direct lattice vectors are
\begin{equation}
\bfa\nd_1=a\,\big(1,0\big)\quad,\quad \bfa\nd_2=
a\big(\frac{1}{2}\,  , \, \frac{\sqrt{3}}{2}\big)\ ,
\end{equation}
with $\bfa\nd_3=\bfa\nd_2-\bfa\nd_1$.  This results in a free energy per site of
\begin{equation}
f(T)=-\lambda(T)+{T\over\pi}\!\int\limits_0^\infty\!\!d\alpha\,\ln\!\bigg(
{\lambda(T)+2+2\cos\alpha\over T}\bigg)\ ,
\end{equation}
with
\begin{equation}
\lambda(T)=\sqrt{4+T^2}-2\ .
\end{equation}

For the $\sqrt{3}\times\sqrt{3}$ structure, the underlying lattice is again triangular, but 
now with a nine element basis (see fig. \ref{kag_rt3}).  The Hamiltonian $\HLT$ is then
purely local, and there is no dispersion.  The density of states per site is found to be
\begin{equation}
g(\ve)=\frac{1}{6}\,\delta(\ve) + \frac{1}{3}\,\delta(\ve-1) +  \frac{1}{3}\,\delta(\ve-3) +  
\frac{1}{6}\,\delta(\ve-4)\ .
\end{equation}
The free energy per site is
\begin{equation}
f(T)=2-u(T) + \frac{1}{6} T\,\ln\!\Bigg({ \big(u^2(T)-4\big)\,\big( u^2(T)-1\big)^2\over T^6}
\Bigg)\ ,
\end{equation}
with $u(T)\equiv\lambda(T)+2$ satisfying
\begin{equation}
{1\over T}={u\,(u^2-3)\over (u^2-1)\,(u^2-4)}\ .
\end{equation}

Our results are plotted in fig. \ref{ferg}.  One finds at all temperatures $T>0$ that 
\begin{equation}
f\nd_{\sqrt{3}\times\sqrt{3}}(T) < f\nd_{\bfQ=0}(T)\ ,
\end{equation}
suggesting that the local correlations should be better described by the
$\sqrt{3}\times\sqrt{3}$ structure.

\section{At the edge}
\label{edge}
With periodic boundary conditions applied, the translationally invariant AKLT states are
nondegenerate.  On systems with a boundary, the AKLT models exhibit completely free
edge states, described by a local spin $S\nd_\rme$ on each edge site which is smaller
than the bulk spin $S$.  The energy is independent of the edge spin configuration, hence
there is a ground state entropy $(2S\nd_\rme+1)\,\kB\ln N_\rme$, where $N\nd_\rme$ is the
number of edge sites.  As one moves away from the AKLT point in the space of Hamiltonians,
the degeneracy is lifted and the edge spins interact.

The existence of weakly interacting $S=\half$
degrees of freedom at the ends of finite $S=1$ Heisenberg chains was first discussed by
Kennedy \cite{KEN90}, who found numerically an isolated quartet of low energy states
for a one-parameter family of $S=1$ antiferromagnetic chains.   These four states are arranged
into a singlet and a triplet, corresponding to the interaction of two $S=\half$ objects.  The spin
quantum number of the ground state alternates with chain length $L$: singlet for even $L$ and triplet
for odd $L$.  The singlet-triplet splitting was found to decay exponentially in $L$ as $\exp(-L/\xi)$, where
$\xi$ is the spin-spin correlation length.   Thus, for long chains, the $S=\half$ objects at the ends are
independent.  Experimental evidence for this picture was adduced from ESR studies of the compound
NENP \cite{HAG90}.

\begin{figure}[!t]
\centering
\includegraphics[width=6.5cm]{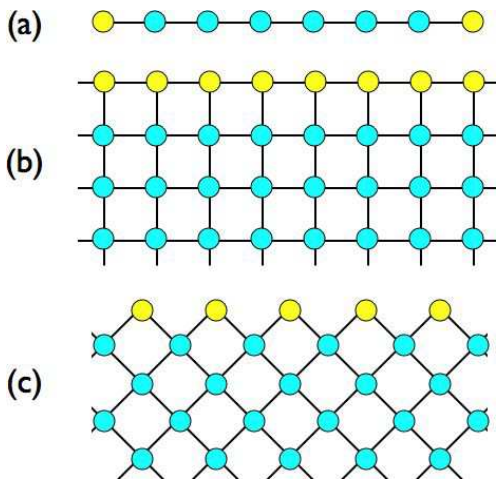}
\caption
{\label{edge_states} (a) Edge states for the linear chain.  (b) $(10)$ edge sites for the square lattice.
(c) $(11)$ edge sites for the square lattice.   For the AKLT states, there is an effective free spin
of length $S_\rme=\half M(z-z_\rme)$ on the edge sites (see text).}
\end{figure}

The situation is depicted in fig. \ref{edge_states}
for the linear chain and for the $(10)$ and $(11)$ edges on the square lattice.  Recall that each
link in the AKLT model supplies one Schwinger boson to each of its termini.  The spin on any site
is given by half the total Schwinger boson occupation: $S=\half(b\yd_\uar b\nd_\uar +
b\yd_\dar b\nd_\dar)$.  Consider first the $M=1$ AKLT state of eqn. \ref{VBS} on the linear chain.
The bulk sites have total boson occupancy $n=2$, hence $S=1$, while the end sites have $n=1$,
hence $S=\half$.  If the end sites are also to have $S=1$, they must each receive an extra Schwinger
boson, of either spin ($\uar$ or $\dar$).  Thus, the end sites are described by an effective $S=\half$
degree of freedom.  Each of these four states is an exact ground state for the AKLT Hamiltonian, 
written as a sum over projection operators for total bond spin $S\nd_{n,n+1}=2$ \cite{AKLT}.

Consider next the square lattice with $M=1$, for which the bulk spin is $S=2$.   For a $(10)$ 
edge, the edge sites are threefold coordinated, and each is `missing' one Schwinger boson.
The freedom in supplying the last Schwinger boson corresponds once again to a $S=\half$
object at each edge site.  Along the $(11)$ edge, the sites are twofold coordinated, and must
each accommodate two extra bosons, corresponding to $S=1$.  The general result for
the edge spin $S\nd_\rme$ is clearly
\begin{equation}
S\nd_\rme=\half M(z-z\nd_\rme)\ ,
\end{equation}
where $z$ and $z\nd_\rme$ are the bulk and edge coordination numbers.  I stress that
the edge spin configurations are completely degenerate at the AKLT point, since all the
internal links are satisfied, \ie\ annihilated by the local projector(s) in the corresponding
AKLT Hamiltonian.  Moving off of the AKLT point, in the direction of the Heisenberg model, the edge
spins will interact.  Based in part on Kennedy's results, I conclude that the $S=\half$ chain
along the $(10)$ edge is antiferromagnetic, while the $S=1$ chain along the $(11)$ edge
is ferromagnetic (since consecutive edge sites are connected through an odd number of bulk sites).

By deriving and analyzing lattice effects in the spin path integral for the Heisenberg model,
\ie\ tunneling processes which have no continuum limit and which do not appear in the
effective nonlinear sigma model, Haldane \cite{HAL88} argued that Heisenberg antiferromagnets
with $2S=0\ {\rm mod}\ 4$ on the square lattice should have nondegenerate bulk ground states.
This result was generalized by Read and Sachdev \cite{RS89}, who, building on an earlier large-$N$
Schwinger boson theory \cite{AA88}, extended Haldane's analysis to SU$(N)$, for $(\bfN,{\bar\bfN}$)
models on bipartite lattices.  This established a connection to the AKLT states, which are 
nondegenerate in the bulk, and which exist only for $S=2M$ on the square lattice.

For the simplex solids, a corresponding result holds.  Recall the $N=3$, $M=1$ model
on the Kagom{\'e} lattice, where each site is in the fully symmetric, six-dimensional \ $\yng(2)$\ \ representation, whose wavefunction $\ket{\!\rmPsi\!}$ is given in eqn. \ref{SSkag}.
Along a $(10)$ edge, as in fig. \ref{kagome}, the edge sites each belong to a single simplex.
Hence, they are each deficient by one Schwinger boson.  The freedom to supply this missing
boson on each edge site is equivalent to having a free edge spin in the fundamental representation
at every site.  Thus, as in the SU(2) AKLT case, the bulk SU($N$) representation is `fractionalized'
at the edge.  The \ $\yng(1)$\ \ objects along the edge are of course noninteracting and degenerate
for the SS projection operator Hamiltonian.  For general SU($N$) models which are in some sense
close to the SS model, these objects will interact.

\section{Conclusions}
I have described here a natural generalization of the AKLT valence bond solid
states for SU$(2)$ quantum antiferromagnets.  The new simplex solid
states $\ket{\rmPsi(\cL\,;\,M)}$ are defined by the application of $N$-site SU$(N)$
singlet creation operators $\cR\yd_\Gamma$ to $N$-site simplices $\Gamma$ of a
particular lattice. For each lattice, a hierarchy of SS states is defined, parameterized
by an integer $M$, which is the number of singlet operators per simplex.  The SS
states admit a coherent state description in terms of ${\rm CP}^{N-1}$ variables, and
using the coherent states, one finds that the equal time correlations in
$\ket{\rmPsi(\cL\,;\,M)}$ are equivalent to the finite temperature correlations of
an associated classical ${\rm CP}^{N-1}$ spin Hamiltonian $\Hcl$, and on the same lattice.
The fictitious temperature is $T=1/M$, and a classical ordering at $T\nd_\rmc$
corresponds to a quantum phase transition as a function of the parameter $M$.
This transition was investigated using a simple mean field approach.
I further argued that for $N>2$ the ordered structure is selected by an
`order by disorder' mechanism, which in the classical model amounts to an entropic
favoring of one among many degenerate $T=0$ structures.  I hope to report on
classical Monte Carlo study of $\Hcl$ on Kagom{\'e} ($N=3$) and pyrochlore
($N=4$) lattices in a future publication; there the coherent state formalism derived here
will be more extensively utilized.  Finally, a kind of `fractionalization' of the bulk
SU$(N)$ representation at the edge was discussed.

\section{Acknowledgments} This work grew out of conversations with Congjun Wu,
to whom I am especially grateful for many several stimulating and useful discussions.
I thank Shivaji Sondhi (who suggested the name ``simplex solid") for reading the
manuscript and for many insightful comments.  I am indebted to Martin Greiter and
Stephan Rachel for a critical reading of the manuscript, and for several helpful suggestions
and corrections.  I also gratefully acknowledge discussions with Eduardo Fradkin.

\section{Appendix I : Properties of SU($N$) coherent states}

\subsection{Definition of SU($N$) coherent states}
Consider the fully symmetric representation of SU($N$) with $p$ boxes in a single row
I call this the $p$-representation, of dimension $N+p-1\choose p$.  Define the state
\begin{equation}
\ket{z\,;\,p}={1\over\sqrt{p!}}\,\Big(z\nd_1\,b\yd_1 + \ldots + z\nd_N\,b\yd_N\Big)^p\,
\ket{0}\ ,
\end{equation}
where $z\in {\rm CP}^{N-1}$ is a complex unit vector, with $z\yd z=1$.
In order to establish some useful properties regarding the SU($N$) coherent states,
it is convenient to consider the unnormalized coherent states
\begin{equation}
\ketB{z\,,\xi}=\exp\big(\xi\, z\nd_\mu\,b\yd_\mu\big)\,\ket{0}=\sum_{p=0}^\infty
{\xi^p\over\sqrt{p!}}\,\ket{z\,;\,p}\ .
\end{equation}
Clearly $\ketB{z\,,\xi}$ is a product of $N$ (unnormalized) coherent states for the
$N$ Schwinger bosons.  One then has
\begin{align}
\braketB{z\,,\xi}{z'\,,\xi'}&= \exp\big(\xibar\xi'\,z\yd z'\big)\nonumber\\
&=\sum_{p=0}^\infty { (\xibar\xi)^p\over p!}\,\braket{z\,;\,p\,}{z'\,;\,p} 
\end{align}
Equating the coefficients of $(\xibar\xi)^p$, one obtains the coherent state overlap,
\begin{equation}
\braket{z\,;\,p}{z'\,;\,p}=\big(z\yd z'\big)^p\ .
\label{CSO}
\end{equation}

\subsection{Resolution of identity}
Define the measure
\begin{equation}
d\mu=\prod_{j=1}^N {d\,{\rm Re}(z_j)\,d\,{\rm Im}(z_j)\over \pi}\,\delta\big(z\yd z-1\big)
\end{equation}
Next, consider the expression
\begin{align}
\cP(\xi,\xibar)&=\int\!d\mu\,\ketB{z\,,\xi}\braB{z\,,\xi}\nonumber\\
&=\sum_{n\nd_1\cdots n\nd_N}
{\big(\xibar\xi\big)^{\sum_j n\nd_j}_{\vphantom{\dagger}}\over\prod_j n\nd_j !}
\ I^N_{n\nd_1\cdots n\nd_N}\,\ket{ \bfn }\ \bra{\bfn}
\nonumber\\
&=\sum_{p=0}^\infty {\big(\xibar\xi\big)^p\over p!}\int\!d\mu\>\ket{z\,;\,p}\>\bra{z\,;\,p}
\label{Peqn}
\end{align}
where $\ket{\bfn}=\ket{n\nd_1,\ldots,n\nd_N}$ and
\begin{equation}
I^N_{n\nd_1\cdots n\nd_N}\equiv\int\limits_0^1\!\!dx\nd_1
\cdots \int\limits_0^1\!\!dx\nd_N\ \delta\bigg(\sum_{j=1}^N x\nd_j-1\bigg)\,
\prod_{j=1}^N x_j^{n\nd_j}\ .
\end{equation}
Here, $x\nd_j=\big|z\nd_j\big|^2$.  If we define $x\nd_j=(1-x\nd_N)\,y\nd_j$ for $j=1,
\ldots,N-1$, then by integrating over $x\nd_N$ one obtains the result
\begin{align}
I^N_{n\nd_1\cdots n\nd_N}&=I^{N-1}_{n\nd_1\cdots n\nd_{N-1}}\cdot
\int\limits_0^1\!\!dx\nd_N\>x_N^{n\nd_N}\,\big(1-x\nd_N)^{N-2+\sum_{j=1}^{N-1}n\nd_j}
\nonumber\\
&={n\nd_N !\ \Big(N-2+\sum_{j=1}^{N-1}n\nd_j\Big)\,!\over \Big(N-1+\sum_{j=1}^{N-1}
n\nd_j  \Big)\,!}\cdot I^{N-1}_{n\nd_1\cdots n\nd_{N-1}}\nonumber\\
&={n\nd_1!\,\cdots\,n\nd_N!\over \big(N-1+n\nd_1+\ldots+n\nd_N\big)\,!}\ .
\end{align}
Thus, equating the coefficient of $\big(\xibar\xi\big)^p$ in eqn \ref{Peqn},
one arrives at the result
\begin{equation}
\bone_p={ (N-1+p)\,!\over p\,!}\int\!d\mu\>\ket{z\,;\,p}\bra{z\,;\,p}\ ,
\end{equation}
where the projector onto the $p$-representation is
\begin{equation}
\bone_p\equiv\sum_{n\nd_1\cdots
n\nd_N}\delta\nd_{p,\sum_j n\nd_j}\,\ket{\bfn}\>\bra{\bfn}\ .
\end{equation}

\subsection{Continuous representation of a state $\ket{\psi}$}
Let us define the state
\begin{align}
\ket\psi&\equiv {1\over\sqrt{p\,!}}\,\psi\big(b\yd_1,\ldots,b\yd_N\big)\ket{0}\nonumber\\
&={1\over\sqrt{p\,!}}{\sum_\bfn}'\,\psi\nd_\bfn\,
{b\yd_1}^{n\nd_1}\cdots{b\yd_N}^{\!\!n\nd_N}\ket{0}\ ,
\end{align}
where $\bfn=\{n\nd_1,\ldots,n\nd_N\}$, and where the prime on the sum reflects
the constraint $\sum_{j=1}^N n_j=p$.  The overlap of $\ket{\psi}$ with the coherent
state $\ket{z\,;\,p}$ is
\begin{align}
\braket{z\,;\,p}{\psi}&={\sum_\bfn}'\psi\nd_\bfn\,\zbar_1^{n\nd_1}
\cdots \zbar_N^{n\nd_N}\\
&= \psi(\zbar\nd_1,\ldots,\zbar\nd_N)\ .
\end{align}

\subsection{Matrix elements of representation-preserving operators}
Next, consider matrix elements of the general representation-preserving operator,
\begin{equation}
{\hat T}\nd_K={\sum_{\bfm,\bfn}}'\ T\nd_{\bfk,\bfl}\ {b\nd_1}^{\!\!\!k\nd_1}\cdots
{b\nd_N}^{\!\!\!\!\!k\nd_N}\>{b\yd_1}^{l\nd_1}\cdots{b\yd_N}^{\!\!l\nd_N}\ .
\label{TKOA}
\end{equation}
Here the prime on the sum indicates a constraint $\sum_{j=1}^N k\nd_j=
\sum_{j=1}^N l\nd_j=K$.
Then
\begin{align}
\expect{\phi}{{\hat T}\nd_K}{\psi}&={1\over p\,!}{\sum_{\bfm,\bfn\atop\bfk,\bfl}}''
T\nd_{\bfk,\bfl}\,\phi^*_\bfm\,\psi\nd_\bfn\\
&\qquad\qquad\times \prod_{j=1}^N \Big[(m\nd_j+k\nd_j\big)!
\ \delta\nd_{m\nd_j+k\nd_j,n\nd_j+l\nd_j}\Big]\ ,\nonumber
\end{align}
where the double prime on the sum indicates constraints on each of the sums
for $\bfm$, $\bfn$, $\bfk$, and $\bfl$.
This may also be computed as an integral over coherent state wavefunctions:
\begin{align}
\int\!\!d\mu\>{\bar\phi}&(z)\,T\nd_K(z,\zbar)\,\psi(\zbar)=
\sum_{\bfm,\bfn\atop\bfk,\bfl}
{\bar\phi}\nd_\bfm\,T\nd_{\bfk,\bfl}\,\psi\nd_\bfn\nonumber\\
&\qquad\qquad\qquad\qquad\qquad\times\prod_{j=1}^N
z_j^{m\nd_j+k\nd_j}\,\zbar_j^{n\nd_j+l\nd_j}\nonumber\\
&={\sum_{\bfm,\bfn\atop\bfk,\bfl}}'' I^N_{\bfm+\bfk}\,
{\bar\phi}\nd_\bfm\,T\nd_{\bfk,\bfl}\,\psi\nd_\bfn\>\prod_{j=1}^N
\delta\nd_{m\nd_j+k\nd_j,n\nd_q+l\nd_n}\nonumber\\
&={p\,!\over\big(N-1+p+K\big)!}\,\expect{\phi}{{\hat T}\nd_K}{\psi}\ .
\end{align}
Thus, the general matrix element may be written
\begin{equation}
\expect{\phi}{{\hat T}\nd_K}{\psi}={\big(N-1+p+K\big)!\over p\,!}
\int\!\!d\mu\ {\bar\phi}(z)\,T\nd_K(z,\zbar)\,\psi(\zbar)\ ,
\end{equation}
where $T\nd_K(z,\zbar)$ is obtained from eqn. \ref{TKOA} by substitution
$b\nd_j\to z\nd_j$ and $b\yd_j\to \zbar\nd_j$.

\section{Appendix II : The local mean field}
With the definition of eqn. \ref{Qeqn}, I first compute
\begin{align}
R\nd_N(m)&\equiv\cQ^{(1)}\wedge\cdots\wedge\cQ^{(N)}\nonumber\\
&=\eps^{\alpha\nd_1\ldots\alpha\nd_N}\,\eps^{\beta\nd_1\ldots\beta\nd_N}\,
\Big({1-m\over N}\,\delta\nd_{\alpha\nd_1\beta\nd_1}+ m\,
\rmP^1_{\!\alpha\nd_1\beta\nd_1}\Big)\nonumber\\
&\qquad\qquad\cdots \Big({1-m\over N}\,\delta\nd_{\alpha\nd_N\beta\nd_N}+ m\,
\rmP^N_{\!\alpha\nd_N\beta\nd_N}\Big)\ ,
\end{align}
where
\begin{equation}
\rmP^\sigma_{\alpha\beta}=\omega^\sigma_\beta\,{\bar\omega}^\sigma_\alpha
\end{equation}
is the projector onto subspace spanned by $\omega^\sigma$.  I now systematically
expand in powers of the projectors and contract over all free indices.  The result is
\begin{align*}
R\nd_N(m)=N!\>\sum_{j=0}^N {m^l\over j!}\,\bigg({1-m\over N}\bigg)^{\!\!N-j}\ .
\end{align*}

The local mean field on a $j=1$ site is given by the expression in eqn. \ref{lmf}.
Expanding the numerator,
\begin{equation}
R\nd_N(m)\,h^N_{\alpha\nd_N\beta\nd_N}=\eps^{\alpha\nd_1\alpha\nd_2
\ldots\alpha\nd_N}\,\eps^{\beta\nd_1\beta\nd_2\ldots\beta\nd_N}\,
\cQ^{(1)}_{\alpha\nd_1\beta\nd_1}\!\!\cdots\cQ^{(N-1)}_{\alpha\nd_{N-1}\beta\nd_{N-1}}\ ,
\end{equation}
in powers of the projectors, the term of order $j$ is
\begin{align}
&(N-j-1)!\,\bigg({1-m\over N}\bigg)^{\!N-j-1}\!m^j\ 
\eps^{\alpha\nd_1\alpha\nd_2\ldots\alpha\nd_N}\,
\eps^{\beta\nd_1\beta\nd_2\ldots\beta\nd_N}\,\nonumber\\
&\qquad\times\!\!\!\!\!\! \sum_{k\nd_1 < \ldots < k\nd_j<N}\!\!\!\!\!
\rmP^{k\nd_1}_{\!\alpha\nd_1\beta\nd_1}\,
\rmP^{k\nd_2}_{\!\alpha\nd_2\beta\nd_2}\cdots
\rmP^{k\nd_{N-1}}_{\!\alpha\nd_{N-1}\beta\nd_{N-1}}
\label{ordj}
\end{align}
Writing
\begin{equation}
\eps^{\alpha\nd_1\alpha\nd_2\ldots\alpha\nd_N}\,
\eps^{\beta\nd_1\beta\nd_2\ldots\beta\nd_N}=
\sum_{\sigma\in\cS\nd_N}{\rm sgn}(\sigma)\>
\delta\nd_{\alpha\nd_1\beta\nd_{\sigma(1)}}\cdots
\delta\nd_{\alpha\nd_N\beta\nd_{\sigma(N)}}\ ,
\end{equation}
we see that once this is inserted into eqn. \ref{ordj}, the only surviving permutations
are the identity, and the $(N-1)$ two-cycles which include index $N$.  All other
permutations result in contractions of indices among orthogonal projectors, and
hence yield zero.  Furthermore using completeness, we have
\begin{equation}
\sum_{i=1}^{N-1} \rmP^{i}_{\!\beta\nd_N\alpha\nd_N}=
\delta\nd_{\beta\nd_N\alpha\nd_N}\!\!-\ \rmP^{(N}_{\!\beta\nd_N\alpha\nd_N}\ .
\end{equation}
Thus,
\begin{align}
&\eps^{\alpha\nd_1\alpha\nd_2\ldots\alpha\nd_N}\,
\eps^{\beta\nd_1\beta\nd_2\ldots\beta\nd_N}
\!\!\!\!\!\! \sum_{k\nd_1 < \ldots < k\nd_j<N}\!\!\!\!\!
\rmP^{k\nd_1}_{\!\alpha\nd_1\beta\nd_1}\,
\rmP^{k\nd_2}_{\!\alpha\nd_2\beta\nd_2}\cdots
\rmP^{k\nd_{N-1}}_{\!\alpha\nd_{N-1}\beta\nd_{N-1}}\nonumber\\
&\quad={N-1 \choose j}\ \delta\nd_{\beta\nd_N\alpha\nd_N} -\!\!\!
\sum_{k\nd_1 < \ldots < k\nd_j<N}
\ \sum_{l=1}^j \rmP^{k\nd_l}_{\!\beta\nd_N\alpha\nd_N}=\nonumber\\
&\quad={N-1\choose j}\ \delta\nd_{\beta\nd_N\alpha\nd_N} -
\ \ {N-2\choose j-1}\,\Big(\delta\nd_{\beta\nd_N\alpha\nd_N}-
\ \ \rmP^{N}_{\!\beta\nd_N\alpha\nd_N}\Big)\bvph\nonumber\\
&\quad={N-2\choose j}\ \delta\nd_{\beta\nd_N\alpha\nd_N}  +
{N-2\choose j-1}\ \rmP^{N}_{\!\beta\nd_N\alpha\nd_N}\ .
\end{align}
From this expression, I obtain the results of eqns. \ref{heqn}, \ref{Aeqn},
\ref{Beqn}, and \ref{Reqn}.


\vfill\eject

\end{document}

%% file: ldefs.tex
  \font\elevenmib=cmmib10 scaled 1095
 \font\tenmib=cmmib10
  \font\eightmib=cmmib10 scaled 800
  \font\sixmib=cmmib10 scaled 667
  \skewchar\elevenmib='177
  \newfam\mibfam
  \def\mib{\fam\mibfam\tenmib}
  \textfont\mibfam=\tenmib
  \scriptfont\mibfam=\eightmib
  \scriptscriptfont\mibfam=\sixmib
  \mathchardef\alpha="710B
  \mathchardef\beta="710C
  \mathchardef\gamma="710D
  \mathchardef\delta="710E
  \mathchardef\epsilon="710F
  \mathchardef\zeta="7110
  \mathchardef\eta="7111
  \mathchardef\theta="7112
  \mathchardef\kappa="7114
  \mathchardef\lambda="7115
  \mathchardef\mu="7116
  \mathchardef\nu="7117
  \mathchardef\xi="7118
  \mathchardef\pi="7119
  \mathchardef\rho="711A
  \mathchardef\sigma="711B
  \mathchardef\tau="711C
  \mathchardef\phi="711E
  \mathchardef\chi="711F
  \mathchardef\psi="7120
  \mathchardef\omega="7121
  \mathchardef\varepsilon="7122
  \mathchardef\vartheta="7123
  \mathchardef\varrho="7125
  \mathchardef\varphi="7127
    \def\physgreek{
    \mathchardef\Gamma="7100
    \mathchardef\Delta="7101
    \mathchardef\Theta="7102
    \mathchardef\Lambda="7103
    \mathchardef\Xi="7104
    \mathchardef\Pi="7105
    \mathchardef\Sigma="7106
    \mathchardef\Upsilon="7107
    \mathchardef\Phi="7108
    \mathchardef\Psi="7109
    \mathchardef\Omega="710A}

\def\etal{{\it et al.\/}}

\def\ie{{\it i.e.\/}}

\def\eg{{\it e.g.\/}}

\def\sss#1{{\scriptscriptstyle #1}}

\def\ssr#1{{\sss{\rm #1}}}

\def\kB{k_{\sss\rmB}}
\def\ONE{{1\kern-0.60em 1}}
\def\dsl{\raise.15ex\hbox{$/$}\kern-.57em\hbox{$\partial$}}
\def\nsl{\raise.15ex\hbox{$/$}\kern-.57em\hbox{$\nabla$}}
\def\id{\raise.72ex\hbox{$-$}\kern-.85em\hbox{$d$}\,}
\def\gtwid{\,{\raise.3ex\hbox{$>$\kern-.75em\lower1ex\hbox{$\sim$}}}\,}
\def\ltwid{\,{\raise.3ex\hbox{$<$\kern-.75em\lower1ex\hbox{$\sim$}}}\,}
\def\undr{\raise.3ex\hbox{$\sim$\kern-.75em\lower1ex\hbox{$|\vec x|\to\infty$}}}

\def\frac#1#2{{\textstyle{#1 \over #2}}}
\def\half{\frac{1}{2}}

\def\({\left (}
\def\){\right )}

\def\uar{\uparrow}
\def\dar{\downarrow}

\def\cC{{\cal C}}

\def\cH{{\cal H}}
\def\cI{{\cal I}}

\def\cL{{\cal L}}

\def\cN{{\cal N}}
\def\cO{{\cal O}}
\def\cP{{\cal P}}
\def\cQ{{\cal Q}}
\def\cR{{\cal R}}
\def\cS{{\cal S}}

\def\bfN{{\mib N}}

\def\bfQ{{\mib Q}}
\def\bfR{{\mib R}}
\def\bfS{{\mib S}}

\def\bfa{{\mib a}}

\def\bfh{{\mib h}}

\def\bfk{{\mib k}}
\def\bfl{{\mib l}}
\def\bfm{{\mib m}}
\def\bfn{{\mib n}}

\def\rbfZ{{\bf Z}}

\def\eps{\epsilon}
\def\ve{\varepsilon}

\def\bfsigma{{\mib\sigma}}

\def\xhi{{\raise.35ex\hbox{$\chi$}}}

\def\rmPsi{{\rm\Psi}}

\def\rmA{{\rm A}}
\def\rmB{{\rm B}}
\def\rmC{{\rm C}}

\def\rmP{{\rm P}}

\def\rmc{{\rm c}}

\def\rme{{\rm e}}

\def\rmt{{\rm t}}

\def\nhat{{\hat{\mib n}}}

\def\ket#1{{\big|\, #1 \big\rangle}}
\def\bra#1{{\big\langle #1\, \big|}}
\def\braket#1#2{{\big\langle #1\, \big|\, #2 \big\rangle}}
\def\expect#1#2#3{{\big\langle #1 \,\big|\, #2\, \big| \,#3 \big\rangle}}
\def\ketB#1{{\,\big| \,#1 \big)\,}}
\def\braB#1{{\,\big( #1 \,\big|\,}}
\def\braketB#1#2{{\,\big( #1 \,\big|\, #2 \big)\,}}

\def\nd{^{\vphantom{\dagger}}}

\def\yd{^\dagger}

\def\vph{\vphantom{\sum_i}}
\def\bvph{\vphantom{\sum_N^N}}

\def\ctnh{\,{\rm ctnh\,}}

\def\and{a^{\phantom\dagger}}

\gdef\journal#1, #2, #3, 1#4#5#6{               
    {\sl #1~}{\bf #2}, #3 (1#4#5#6)}            

%% file: simplex_solid.bbl
\begin{references}
\bibitem{FOOT} It is worth remarking that the AKLT states themselves do not require a bipartite lattice,
and indeed can be defined on any lattice.  The corresponding classical model, given in eqn.
\ref{AKLTcl}, is then frustrated.
\bibitem{disc} By taking linear combinations $\ket{\rmPsi\nd_\bigtriangleup}\pm
\ket{\rmPsi\nd_\bigtriangledown}$, one can construct $\cI$ eigenstates with
eigenvalues $\pm 1$.  A similar state of affairs holds for the MG model, where
ground states of well-defined crystal momentum $k=0$ and $k=\pi$ may be
constructed from the degenerate broken symmetry wavefunctions.
\end{references}
